# A Physics-embedded Deep Learning Framework for Cloth Simulation

Zhiwei Zhao[1]

[1] UM-SJTU Joint Institute, Shanghai Jiao Tong University, Shanghai, China

zzw483@sjtu.edu.cn

**Abstract.** Delicate cloth simulations have long been desired in computer graphics. Various methods were proposed to improve engaged force interactions, collision handling, and numerical integrations. Deep learning has the potential to achieve fast and real-time simulation, but common neural network (NN) structures often demand many parameters to capture cloth dynamics. This paper proposes a physics-embedded learning framework that directly encodes physical features of cloth simulation. The convolutional neural network is used to represent spatial correlations of the mass-spring system, after which three branches are designed to learn linear, nonlinear, and time derivate features of cloth physics. The framework can also integrate with other external forces and collision handling through either traditional simulators or sub neural networks. The model is tested across different cloth animation cases, without training with new data. Agreement with baselines and predictive realism successfully validate its generalization ability. Inference efficiency of the proposed model also defeats traditional physics simulation. This framework is also designed to easily integrate with other visual refinement techniques like wrinkle carving, which leaves significant chances to incorporate prevailing macing learning techniques in 3D cloth amination.

**Keywords:** Computer Graphics, Cloth Simulation, Machine Learning, Physics-based Animation.

## 1. Introduction

Cloth is an attractive research topic in computer animation due to its ubiquity in the real world. Cloth simulation plays a crucial role in character dressing, scene decoration, and costume design. The surge in computing power has empowered physics-based simulation (PBS) during past decades, which achieved remarkable levels of visual realism. The formulated physics, however, often relies on models (e.g., position-based dynamics or mass-spring system) that might contain underexplored treatments (e.g., bending model and locking issue). Furthermore, accurate and stable simulation also demands high computational costs, prohibiting PBS from real-time, large-time-step, or large-scale applications. The recent triumph of deep learning and advancements in GPU hardware have renewed machine learning (ML) techniques for many scientific fields. In computer graphics (CG), PBS indeed has existing potential to integrate with ML methods to achieve high simulation accuracy with improved computational efficiency.

To animate the cloth, all related forces need to be computed into nodal velocity. The spatiotemporal evolution of cloth objects is then described by a partial differential equation (PDE). Engaged time integration can be categorized into two types, explicit and implicit methods. Depending on practical demands, they are both widely applied. Blender [1], an open-source software, has implemented both methods for mass-spring simulation. For cases when implicit integration is necessary, researchers have proposed many acceleration and simplification strategies. Among them, Conjugate Gradient [2] and Chebyshev Acceleration [3] can be used to optimize the equation-solving process. Constrained Dynamics [4] introduce dual variables and make the PDE system stable even under highly stiff cloth settings. Projective Dynamics (PD) [5] uses an assumed constant Hessian matrix to simplify the iterative process of the Newton-Raphson method. To a further degree, Position-Based Dynamics (PBD) [6] discards solving PDE equations but iteratively updates mass point positions to make the spring system overall minimally deformed. If external forces are exerted, a position-based energy is defined and meant to be minimized. PBD is widely acknowledged as an efficient and stable method to deal with many kinds of temporal evolution, not limited by cloth but also found in fluid and solid simulation. It is then widely used in commercial software such as Houdini [7] and Marvelous Designer [8] for fast cloth simulation.

All these methods make efforts to improve simulation efficiency while maintaining visual realism of cloth dynamics. However, PBS-type methods can not yet satisfy the demand for high-quality and real-time cloth animation. This is due to expensive costs of massive refined meshes, complex treatments of boundary and collision handling, and stiffness of implicit time integration, all of which are inevitable in PBS strategies. In modern industry, real-time cloth simulations are achieving increasingly significant roles in many applications. Such can be found in 3D computer games, augmented reality(AR), virtual live streaming, simultaneous digital interaction, etc. Therefore, efficient and accurate cloth simulation algorithms remain challenging in computer graphics, whence this paper attempts to fill the gap by leveraging advanced deep learning (DL) techniques and an innovative NN structure.

This work contributes to adopting a physics-encoded neural network to learn cloth simulation. The author proposes a physics-embedded DL framework to generalize cloth simulation and is able to integrate with plugged features and constraint handling, as well as merge other networks for additional cloth refinement. This model bypasses the heavy computation of physics-based cloth simulation, which enables good performance for real-time applications. Meanwhile, it relies less on the abundance of training data compared with existing ML methods.

## 2. Related work

With deep learning achieving unprecedented attention, it has widely been utilized in physics applications. PDE-Net [9] was proposed to study internal connections between spatial operators (e.g. gradient, divergence, and Laplacian) and convolutional Neural network (CNN), including the order of spatial discretization and kernel size of convolutional filter. It verifies mathematically the efficacy of learning a PDE system through ML strategies with designed networks. Physics-informed neural network (PINN) [10,11] marked a transformative adaption of deep learning in physics simulation and leaves lasting implications for the field. PINN specially defines a physic loss given by known governing equations and physical constraints. Together with traditional data loss, the training process is both physics-driven and data-driven, combining unsupervised and supervised features respectively. Therefore, it can benefit from both manners and thus rely less on data abundance and sampling efficiency, while staying critical of provided physical laws. In recent PeRCNN [12], researchers implemented recurrent neural networks to describe a dynamically evolving PDE system. It successfully uses recurrent structure to learn a temporally evolving system, given that time derivatives resemble the concept of hidden layers. Such a network is applicable for predicting the dynamic behavior under certain physics principles. In computer graphics, machine-learning strategies have also attracted wide attention. Such can be found in learning automatic control to achieve designated fluid-rigid interaction through deep reinforcement learning [13,14], adjusting implicit actuators for desired positions and gestures of the differentiable body leveraging gradient back-propagation [15,16]. Except for simulation, deep learning techniques are also used in real-time rendering [17] and text-based modeling [18]. Though physical losses have been utilized in learning cloth animation, physics-encoded neural networks are rarely studied in CG.

In the specific field of cloth simulation, learning-engaged techniques have been widely studied by researchers. Inverse problems are one typical topic that naturally benefits from machine learning and gradient backpropagation. Cloth behaviors are generally induced by physical forces and collision. Liang et al. [19] have explicitly derived gradients of a constraint efficiently handling non-penetration collision. DiffCloth [16] investigates differentiable cloth simulation with dry frictional contact. Despite learning control of cloth behaviors, various types of deep neural networks are also applied to accelerate predictions of cloth animation. Bertiche et al. [20] proposed a recurrent encoder-decoder network to achieve unsupervised deep learning for body-motion-correlated cloth subspace, which is solely realized by physical losses. Differently, MeshGraphNetRP [21] employs data-driven cloth simulation through a recurrent graphic neural network. Except for physically simulated cloth, DeepWrinkles [22] developed a conditional Generative Adversarial Network (cGAN) with U-Net from captured data. It can generate dressed 3D clothes with wrinkle-enhanced normal maps. In recent PeRCNN [12], physics is encoded into network structure to learn general PDE systems. Inspired by such ideas, this work builds a CNN-comprised neural network that can learn from both physical loss and ground truth data.

## 3. Methodology

This section briefly reviews several proposed methods of detail refinement for photorealistic cloth simulation. Then different force components are formulated, which are necessary to perform visual-realistic animation. With the benefit of physics-embedded neural network structure, a deep learning framework is introduced to efficiently learn dynamic features of cloth simulation.

*3.1. Physics-based cloth simulation*

Mass-spring system is commonly adopted to model cloth, which can be found in many CG software. Pieces of fabric are described by mass points concatenated with weightless springs. Though springs do not possess potential energy, they tend to minimize their total elastic energy. Internal forces of cloth are primarily realized by these springs under extension or compression. Together with external forces acting on mass points (e.g. gravity and pressure), the whole system can imitate local impulses on fabric segments. Besides dry friction often processed by Signorini-Coulomb law [23], bending is also a significant factor contributing to the overall realism. Mass points resemble hinge joints, which are not resilient to folding, and springs cannot afford to bend. To tackle this issue, the Dihedral Angle [24] and Quadratic Bending [25] were models proposed to enable cloth with bending effects. Alternatively, a node in the mass-spring system can simply connect with a farther node to prevent folding with neighbor nodes.

Details of wrinkles and folds of cloth and garments depend on mass point resolution. If under large deformation, the nonlinear range of elastic behavior and even the plastic behavior of springs should also be considered. With all extra models of bending, damping, and uniform pressure, the mass-spring system can well simulate a garment deforming and wigging. By adjusting relevant material properties, the cloth can take on different levels of folds and wrinkles, falling by gravity, hold-up by friction and air drag, and blown by wind. Cloth animation is continuously generated via either type of time integration. The explicit one, which relates force components with the former step, is easy to implement and computationally fast. The implicit one, which relates force components with the latter step, is stable and accountable to large time steps. In this work, internal spring forces exist in the 13-point stencil for every mass point. Each node would have connections with 8 surrounding neighbors in the manner of 9-point stencil and 4 farther neighbors in cardinal directions. The skip connection gives an easy way to realize bending effects, and this design also guarantees local isotropy by assigning equal weights in principle directions. Equation (1) formulates the elastic spring force for nodal points:

$$\boldsymbol{f}_{E,i} = -\sum_j E \frac{\boldsymbol{x}_{ij}}{|\boldsymbol{x}_{ij}|} (|\boldsymbol{x}_{ij}| - L_{ij}), \tag{1}$$

where $\boldsymbol{f}_{E,i}$ stands for the total elastic force of point $i$, and $j$ is the neighbor point index. Their relative displacement is denoted by $\boldsymbol{x}_{ij}$, with $|\boldsymbol{x}_{ij}|$ being its magnitude. $E$ is Young's modulus of the spring, and

$L_{ij}$ is the original length. Elastic energy can be recovered periodically for ideal cases. In the real world, however, a spring cannot oscillate forever even without external tension. This is due to the existence of damping, which dissipates elastic energy into low-level thermal energy and releases it to surroundings. Damping forces are generated to decline relative motion of two endpoints of each spring, which is formulated by

$$\boldsymbol{f}_{\mu,i} = -\sum_j \mu(\boldsymbol{v}_{ij} \cdot \frac{x_{ij}}{|x_{ij}|}) \frac{x_{ij}}{|x_{ij}|}, \qquad (2)$$

where $\boldsymbol{f}_{\mu,i}$ stands for total damping force of point $i$, related to damping coefficient $\mu$ and relative velocity $\boldsymbol{v}_{ij}$. Cloth and garments can interact with wind and pressure frequently in reality, such as hanging laundry or blowing up airbags. Pressure forces are usually uniform in magnitude and always act orthogonally on the cloth surface. If evenly distributed on vertices, the force on mass points can be calculated through cross product as

$$\boldsymbol{f}_{p,i} = \sum_{j_1,j_2} p(x_{ij_1} \times x_{ij_2}). \qquad (3)$$

Unlike neighbors in the 13-point stencil, nodal pressure force $\boldsymbol{f}_{p,i}$ only takes account of the 4 nearest points by which triangular faces are formed. They are permutated in 4 pairs in a clockwise direction, each denoted by $(j_1, j_2)$. Within a pair, $j_2$ can be viewed as rotating $j_1$ by a quarter of the circle in orthogonal axes. $x_{ij}$ denotes relative displacement with the central point, and $p$ is the local pressure. Other external forces may include gravity and air drag. The former is given by nodal mass $m_i$ and a constant acceleration $\boldsymbol{g}$. The latter, which is quantified by the drag coefficient $d$, gives a contrary impulse to nodal velocity $\boldsymbol{v}_i$:

$$\boldsymbol{f}_{ext,i} = m_i \boldsymbol{g} - d\boldsymbol{v}_i. \qquad (4)$$

Depending on whether part of the cloth is fixed or moves along with other objects, Dirichlet or Neumann boundary conditions can be applied. Through iterations of internal interaction (i.e., elastic and damping forces), such boundary constraints can propagate through the whole spring. To reduce animation artifacts, collision handling is another vital point to deal with. To detect either cloth-object collision or cloth self-collision, broad-phase spatial partition or bounding section hierarchy is first performed. Depending on simulation time step or accuracy requirement, a continuous detection or discrete detection is performed locally for potentially colliding nodes. To prevent or eliminate collision, various methods have been proposed, such as Impact Zone Optimization [26] and Untangling Cloth [27]. In this paper, the author mainly employs intersection elimination for cloth-object collision but leaves cloth self-collision for future work. Towards deep learning efficacy in cloth simulation, detail treatments can improve visual realism, but a simplification would not weaken general feasibility of such work. Denoting $\mathscr{C}$ as all treatments of boundary conditions (BCs), collisions, and other constraints. Through a semi-implicit integration, the dynamic process of cloth animation can be formulated as equation (5), where $k$ stands for time iteration covering a time step $\delta t$, $\boldsymbol{f}$ is the total force and $M^{-1}$ is the mass matrix.

$$\boldsymbol{v}^{(k+1)} = \mathscr{C}(\boldsymbol{v}^{(k)} + \sum \boldsymbol{f}^{(k)} M^{-1} \delta t), \quad \boldsymbol{x}^{(k+1)} = \boldsymbol{x}^{(k)} + \boldsymbol{v}^{(k+1)} \delta t \qquad (5)$$

*3.2. Deep learning framework for cloth simulation*

According to the above, comprehensive physical information is presented in delicate cloth animation. In computer graphics, studies of ML methods for cloth simulation have applied deep fully connected layers [20], recurrent graphic neural networks [21], etc. Previous machine learning applications, usually, adopt classical DL models which are costly to represent subtle cloth dynamics. Nonetheless, a directly physics-characterized network may serve a parameter-efficient yet inference-accurate learning process. This work tends to investigate such NN structures that directly encode physical features of cloth behavior, which is supposed to increase learning efficiency and inference accuracy.

One principle of designing the neural network is to consider how well the structure can represent hind features of the engaged problem. Convolutional networks have achieved broad success in capturing spatial and frequency correlations, which are significant in digging for information from images or audios. Its efficacy in representing spatial operators is also verified in previous research [9]. Simulating cloth is generally the marching problem of PDE system. Thus, this work designs a CNN-included neural network with additional branches to engage physical features for cloth animation. As described in section 3.1, internal forces occur in a 13-point stencil. A convolutional network is first acted upon $x$ and its derivative $v$, extracting two-point correlations of total 12 pairs each with the central point. Two assumptions can be made in this process, efficiently reducing the number of trainable parameters. The first is that physical properties are isotropic, thus the CNN can be made identical for three dimensions, resulting in 2D convolutional kernels but not 3D. The second is that as each channel features the correlation with one neighbor, corresponding filter may be predetermined with remaining parameters frozen during training. Additionally, dilation kernels can be utilized as the 13-point stencil is sparse. After channel extraction, the neuron would diverge into three different branches. A linear branch simply goes through a linear layer. A nonlinear branch goes through an activation function and is then no-bias linearized. A derivative branch multiplies nonlinearized displacement features with linearized velocity features. Bias is performed only once at linear branch to characterize external forces. If such characteristics are homogeneous or steady, parameter size can be further reduced. At last, all channel information is summed to retrieve 3D force representations. Equations (6) and (7) formulate manipulations of the architecture, where the order of nonlinear and derivative operations can be tuned by trying different activation functions.

$$\begin{cases} \widehat{\boldsymbol{\varphi}}_C^{(k)} = \mathcal{K}(w_k; c)(x^{(k)}), \\ \widehat{\boldsymbol{\xi}}_C^{(k)} = \mathcal{K}(w_k; c)(v^{(k)}) \end{cases} \quad (6)$$

$$\begin{cases} \widehat{\mathcal{L}}^{(k)} = \sum_{c=1}^{Nc}(w_L \cdot \widehat{\boldsymbol{\varphi}}_C^{(k)} + b_L), \\ \widehat{\mathcal{N}}^{(k)} = \sum_{c=1}^{Nc}\left[w_N \cdot \sigma(\widehat{\boldsymbol{\varphi}}_C^{(k)})\right], \\ \widehat{\mathcal{D}}^{(k)} = \sum_{c=1}^{Nc}\left[w_D \cdot \widehat{\boldsymbol{\xi}}_C^{(k)} \otimes \sigma(\widehat{\boldsymbol{\varphi}}_C^{(k)})\right] \end{cases} \quad (7)$$

Here $\widehat{\boldsymbol{\varphi}}_C$ denotes extracted channel features from $x$, and $\widehat{\boldsymbol{\xi}}_C$ is from $v$. $\mathcal{K}(w_k; c)$ is the convolutional kernel, with parameter weights $w_k$ and designated by channel index $c$. $\widehat{\mathcal{L}}$ denotes linear branch features, $\widehat{\mathcal{N}}$ for nonlinear ones and $\widehat{\mathcal{D}}$ for derivative branch. The weights and bias of three branches are denoted by $w$ and $b$ with corresponding subscripts. $\sigma$ is the activation function, $\otimes$ stands for derivative-correlated multiplication. Figure 1 illustrates this neural network structure, where the activation function is chosen as the inverse square root unit (ISRU), which best performs the cloth simulation under current configurations.

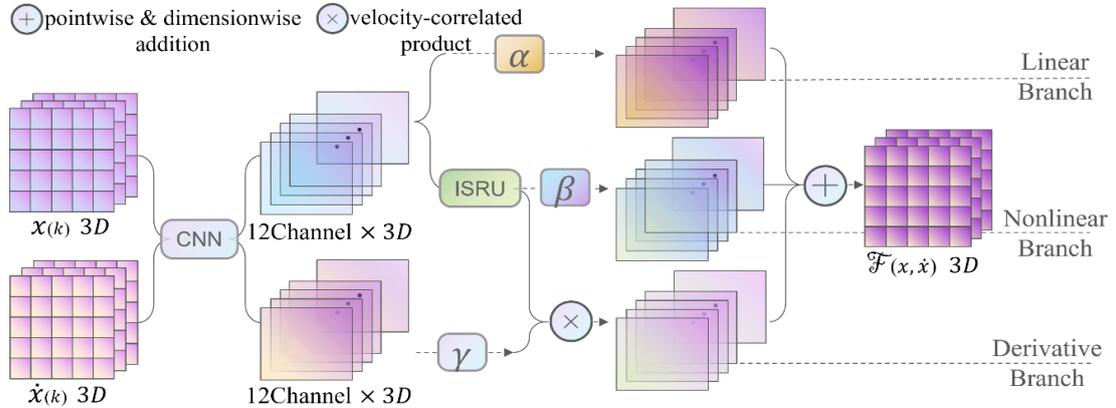

**Figure 1.** Structure of the neural network cell for deriving simulation forces.

As formulated by equation (5), the successive process includes Forward Euler difference to obtain momentum change to the next iteration, as well as cooperating with BCs, collision, and other constraints to update mass point positions to the next frame. Among them, various BC types are hard-encoded to the framework, which is selectable during the runtime. Collison detection and handling are seamlessly integrated before updating the nodal positions. The designed architecture is meant to preserve cloth simulation physics, from velocity computation and position updation. Figure 2 shows the schematic diagram of the framework. Intriguingly, this framework can also be coupled with additional forces (e.g., ghost force and dry friction contact) by traditional PBS, as well as plugged functions by other sub neural networks (e.g., ML for winkle enhancement and collision refinement).

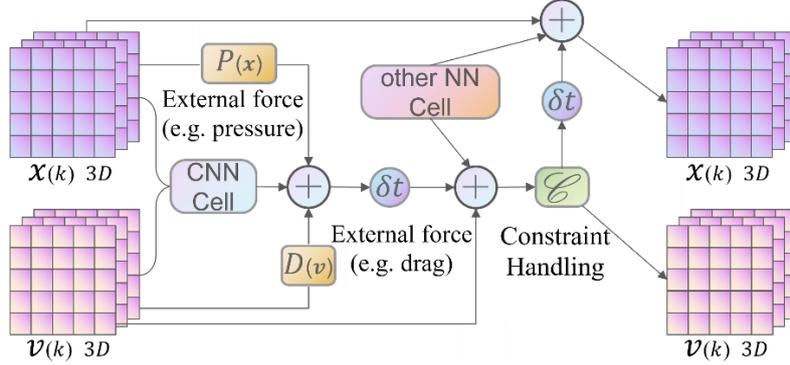

**Figure 2.** Schematic diagram of proposed cloth-simulation framework
(Zoom-in view of CNN Cell is in Figure 1).

*3.2.1. Training process*
To train network parameters, two kinds of losses can be optimized: the physics loss, which is the error between model-predicted impulses $\mathcal{M}$ and equation-solved force integrations, and the data loss, which is the error between network-updated data $[Nx, Nv]$ and ground-truth data. Equation (8) gives the loss expression, where $\alpha$ calculates the weighted average from two loss types and $\langle . \rangle$ stands for mean square error within a batch. Through gradient backpropagation, $w$ and $b$ can be optimized to minimize the loss.

$$\mathcal{L}oss(w,b) = \alpha \langle \mathcal{M}\left(w,b;\begin{bmatrix}x^{(k)}\\v^{(k)}\end{bmatrix}\right) - f^{(k)}\delta_t \rangle$$
$$+ (1-\alpha) \langle \begin{bmatrix}Nx\\Nv\end{bmatrix}\left(w,b;\begin{bmatrix}x^{(k)}\\v^{(k)}\end{bmatrix}\right) - \begin{bmatrix}x^{(k+1)}\\v^{(k+1)}\end{bmatrix} \rangle \tag{8}$$

To achieve the best efficiency of training, the scales and signs of branch parameters, as denoted in figure 1, are predetermined by Differential Evolution [30]. By such means, all weights and biases can be assigned the same learning rate, which gradually slows down during the training. This helps the neural network to optimize parameters much easier. Meanwhile, as an optional post-process, learning schedulers and optimizers can be updated to fine-tune the model, depending on the steepness and oscillation of the loss track.

*3.2.2. Evaluation case setup*
The performance of the proposed model is evaluated on a set of 3D benchmarks. The first is cloth falling by gravity with edges pinned stationary. The second is the same cloth blown up by huge pressure from the downside. PBS data for these two scenes are provided as training sets, with which corresponding inference results can best verify model accuracy. To provide additional test sets, two extra scenes are compared by proposed NN framework and PBS. The third scene is a swinging cloth pinned at one corner, where the breeze is modeled by small pressure forces. The fourth scene is cloth dropped on a ball that can absorb bouncing energy and generate friction preventing relative motion. These two scenes have modified the boundary conditions and pressure magnitudes from training data. Besides, new collision handling and external friction forces are imposed. They can adequately evaluate the model generalization ability and meanwhile test the framework efficacy for various applications of cloth simulations.

## 4. Result

*4.1. Training*

The physics simulation is implemented with TaiChi [28] in Python, while the deep learning framework is implemented with PyTorch [29]. The simulated cloth contains $100 \times 100$ mass points which in total form around $60k$ springs. Ground truth data is prepared up to 5000 steps for continuous animation in each example. Physics loss is computed in the runtime to save VRAM as well as utilize GPU parallelism. Training settings are listed in table 1, where stages including pre and post processes of NN optimization are concluded.

**Table 1.** Data preparation and training settings for three stages.

| Training Process | | Pre-process | |
|---|---|---|---|
| Batch Size | 256/5000 | Batch Preparation | Random sample |
| Epoch | 5000+ | Cond. Parallelism | Boolean tensor |
| Initialization Method | Uniform | Parameter Popul. | $6^6$ |
| Optimizer | Adam | Post-process | |
| Scheduler | Step | Scheduler | Cosine Annealing |
| Start Learning Rate | 1e-2 | Start Learning Rate | 1e-3 |

The network is trained with two sets of data, one of which is simulated with super-pressure acting on single side of the cloth and the other is without. If pressure forces are exerted, dynamics of the cloth animation can behave more complicatedly, including enhanced oscillation and larger deformation. Especially, when the pressure is in the counter direction of gravity, the network would be harder trained to recognize gravitational features. Figure 3 demonstrates the training process of parameters under two data sets. The one without pressure is found to converge more easily as the learning rate stabilizes. In either training case, larger steps are initially taken to locate the optimal zone and then reduced to search for a loss-minimized point. Two subplots in figure 3 exhibit the same initial overcorrections and gradual smoothness. For pressure-free training, the process can be continued without the need to update schedulers. Each training takes around 15 hours with an Nvidia RTX2080Ti (22GB) GPU and the average VRAM usage is about 18 GB with proper memory optimizations.

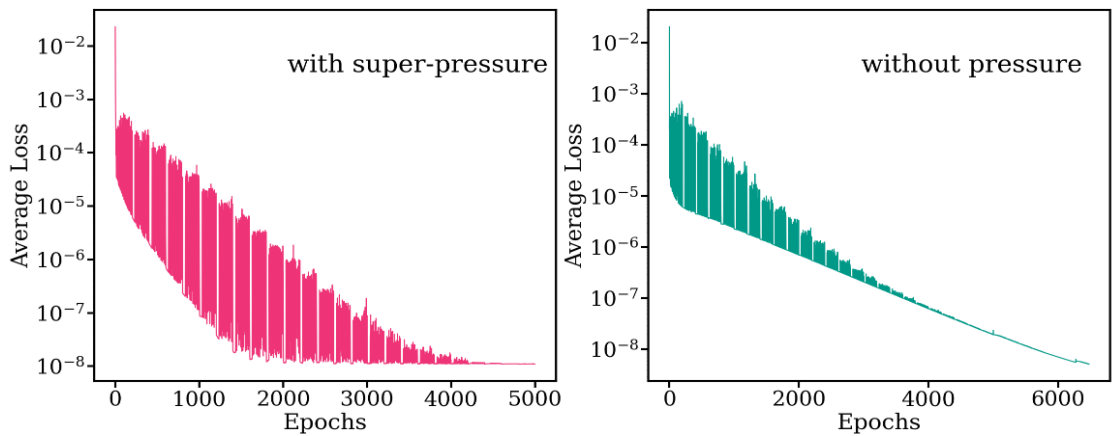

**Figure 3.** Training processes on data from cloth simulations.

*4.2. Evaluation*

The model accuracy is validated in aforementioned animation setups: cloth falling, cloth folding on ball, cloth blown up, and cloth hanging. Figure 4-6 show results in the last three scenes. In each scene, both snapshots are taken at the same frame, where PBS result is treated as the baseline for NN prediction. Note that only in blown-up case the same PBS data is used training the NN model. The other two demonstrations use the same model to predict results under new configurations without retraining.

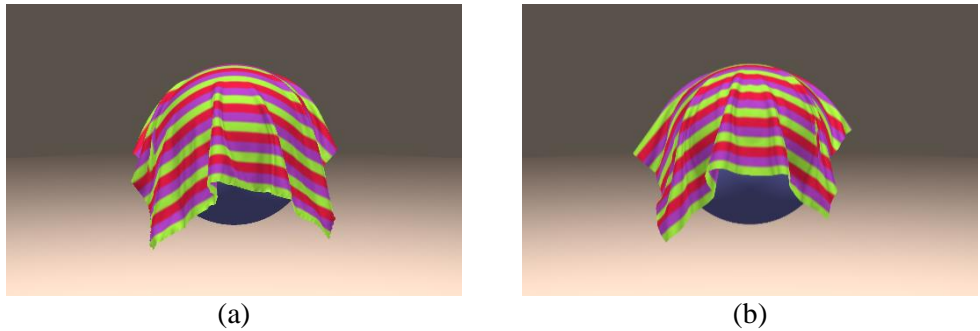

(a)          (b)

**Figure 4.** Cloth folding on ball: (a) NN framework; (b) PBS.

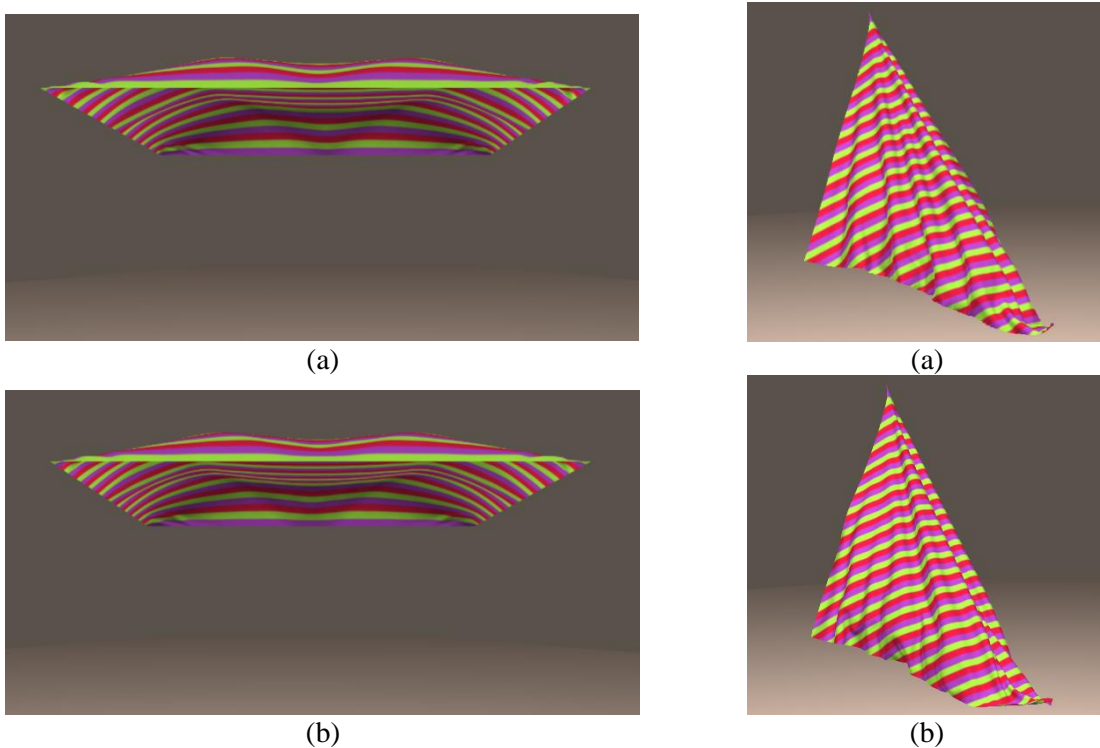

**Figure 5.** Blown-up cloth: (a) this work's NN framework; (b) physics-based simulation (PBS).

**Figure 6.** Hanging cloth: (a) NN framework; (b) PBS.

From figure 1, the inferred result matches exactly the reference, which is no exception as the loss has sufficiently decreased after the learning process. In test sets, the framework keeps excellent performance as seen in figure 5-6. Some tendencies can be concluded from the comparison: gravity might be learned less ideally as the training set has stricter boundary fixation, and the damping effect looks stronger in the model as in figure 5. The folding case takes more dynamic movements related to gravity, so the cloth location falling on ball differs from the baseline. Except for visual effects, the model performance is also tested on computational speed as a common merit owned by deep learning methods. Prediction errors and computing speed comparison are summarized in table 2.

Table 2. Model prediction errors and computational budget.

|  | NN framework | PBS (CPU backend) |
| --- | --- | --- |
| Error (no pressure) | 0.213% | 0% |
| Error (pressure) | 0.477% | 0% |
| Error (cross test[a]) | 0.518% | 0% |
| Computation speed | 232 step/sec | 201 steps/sec |

[a] Model trained without pressure is tested with pressure baseline.

Another merit of the proposed NN model is that, through pattern analysis of the cloth motion, the defect of any dynamic representation can be further optimized with remaining branch parameters frozen. This efficiently eases post training process. For instance, if the cloth cannot fall on the ball at the exact position compared with PBS, the linear branch is to be further fine-tuned. Through different test cases without retraining, performance and generalization ability of the proposed model are proved.

## 5. Conclusions

This paper proposes a physics-embedded deep-learning framework to realize cloth animation. The developed neural network directly encodes physics features of PBS cloth simulation. The structure uses a CNN cell to represent spatial correlation of the mass-spring system, which learns internal interactions as elastic force and damping force. Air drag and constant impulses like gravitation are also informed by a multi-branch design. External forces like pressure, as well as constraints of boundary conditions or object collision, are hard-encoded to the framework to achieve generalization ability. Its performances are validated by comparison with problems outside the scope of training set configurations. Learned parameters can be transferred to handle other test cases within this model, which achieves satisfactory predictions. Meanwhile, the neural network's computational efficiency is improved with GPU from traditional PBS with CPU parallelism.

In future works, functional integrability of the proposed framework will be further investigated. For instance, cloth self-collision can be encoded to eliminate penetration artifacts, and sub-networks of wrinkle enhancement can be added to improve visual realism with low mesh resolution. For the CNN cell, more nonlinear layers and concentrated structures will be studied for larger-step and fully implicit time integrations, both of which are practically useful for fast and stable simulations. For the framework, recurrent structure (i.e., RNN) will be realized to process time-dependency more powerfully in cloth animations.